\documentclass[a4paper,11pt]{article}

\usepackage{float}
\usepackage{pos}
\usepackage{lipsum} %package to generate placeholder text in the following
\usepackage[mathlines]{lineno}
\title{Updated Earth Tomography Using Atmospheric Neutrinos at IceCube}

\ShortTitle{Updated Earth Tomography Using Atmospheric Neutrinos at IceCube}

\author{The IceCube Collaboration \\{\normalsize \normalfont(a complete list of authors can be found at the end of the proceedings)}\\}

\emailAdd{alexwen@icecube.wisc.edu}

\abstract{

The IceCube Neutrino Observatory has observed a sample of high purity, primarily atmospheric, muon neutrino events over 11 years from all directions below the horizon, spanning the energy range 500 GeV to 100 TeV. 
While this sample was initially used for an eV-scale sterile neutrino search, its purity and spanned parameter space can also be used to perform an earth tomography. 
This flux of neutrinos traverses the earth and is attenuated in varying amounts depending on the energy and traversed column density of the event. 
By parameterizing the earth as multiple constant-density shells, IceCube can measure the upgoing neutrino flux as a function of the declination, yielding an inference of the density of each shell. 
In this talk, the latest sensitivities of this analysis and comparisons with the previous measurement are presented. 
In addition, the analysis procedure, details about the data sample, and systematic effects are also explained. This analysis is one of the latest, weak-force driven, non-gravitational, measurements of the earth’s density and mass.
\vspace{4mm}

{\bfseries Corresponding author:}
Alex Y. Wen$^{1*}$\\

{$^{1}$ \itshape Laboratory for Particle Physics and Cosmology and Department of Physics, Harvard University}\\
$^*$ Presenter
}

\ConferenceLogo{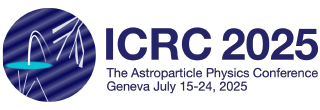}

\FullConference{39th International Cosmic Ray Conference (ICRC2025)\\
 15–24 July 2025\\
Geneva, Switzerland\\}

\begin{document}

\maketitle

\section{Introduction}\label{sec:introduction}

The IceCube neutrino observatory~\cite{Aartsen:2016nxy} is a cubic-kilometer scale neutrino telescope situated in ice at the South Pole.
IceCube can detect astrophysical and atmospheric neutrinos, the latter of which are neutrinos produced via cosmic-ray showers in the Earth's atmosphere.
When these neutrinos interact, via charged current deep inelastic scattering, they produce a charged lepton; for the muon flavored neutrinos, muons are produced, which leave long, characteristic, energy deposits called tracks, in IceCube. 
The key information collected from each neutrino track is the muon's energy and zenith angle. 
The neutrino energy, $E_\nu$, is inferred from the measured muon energy, and can be reconstructed via machine learning methods, like in Ref.~\cite{IceCubeCollaboration:2024dxk}.
The zenith angle, $\theta$, is defined as the angle between the upward-pointing normal vector at IceCube's location, and the vector which points at the source of the neutrino, as reconstructed by the detector. 
A neutrino in the range $-1 < \cos\theta < 0$ is known as upgoing, since these directions correspond to neutrinos which originate from below the horizon. 
In the particular event selection described in detail in Ref.~\cite{IceCubeCollaboration:2024dxk}, IceCube has measured 368,071 primarily atmospheric neutrino tracks in the upgoing direction, over the energy range 500 GeV to 100 TeV in the time period of 10.6 years from 2011-2022.
These tracks can be classified as starting (vertex contained within the detector), or throughgoing (vertex outside the detector). 
More information about this event selection, which was first used for a sterile neutrino search, can be found at Refs.~\cite{IceCubeCollaboration:2024dxk,IceCubeCollaboration:2024nle}; we will use these events to perform our Earth tomography. 

One modern understanding of the Earth's density profile is the PREM model~\cite{PREM:1981297}, which was derived from seismological measurements.
Atmospheric neutrinos can also be used to make such a measurement, offering a unique opportunity to probe the Earth's interior.
Our sample of upgoing neutrinos has traversed the Earth at various angles, corresponding to various column densities (the density integrated along the neutrino's path through the Earth). 
At around 10 TeV, Earth-scale column densities have a non-negligible attenuation effect on the neutrino flux.
By analyzing the number of neutrinos that are observed at a certain point in the space of energy and zenith angle, we can measure the column density in that direction.
Furthermore, by assuming that the Earth is a sphere with concentric layers, we may infer its density profile, mass, and moment of inertia.
Using neutrinos, we will perform a primarily weak force-based measurement, complementing measurements using gravity and electromagnetism.

%At lower energies, a similar measurement of density can be made by considering the matter-induced oscillations of neutrinos; for more information on that analysis, see~\ref{placeholder}. 
% If we compare the observation with an expectation using a tunable radial profile of the Earth's density, we can make an inference of the Earth's density profile and therefore its mass and moment of inertia.
Ref.~\cite{Donini:2018tsg} performed such a measurement using only a year of IceCube data. 
While it had large uncertainties, this measurement was a successful proof of concept that neutrinos can be used to perform a tomography of the Earth. 
Compared to that work, we have greatly enhanced statistics, a more updated treatment of systematic uncertainties, and improved energy reconstruction.
In this work, we present the latest sensitivities of an IceCube analysis to measure this density profile. 

\section{Methods}\label{sec:methods}

\begin{figure}[t!]
\centering
\includegraphics[width=0.9\linewidth]{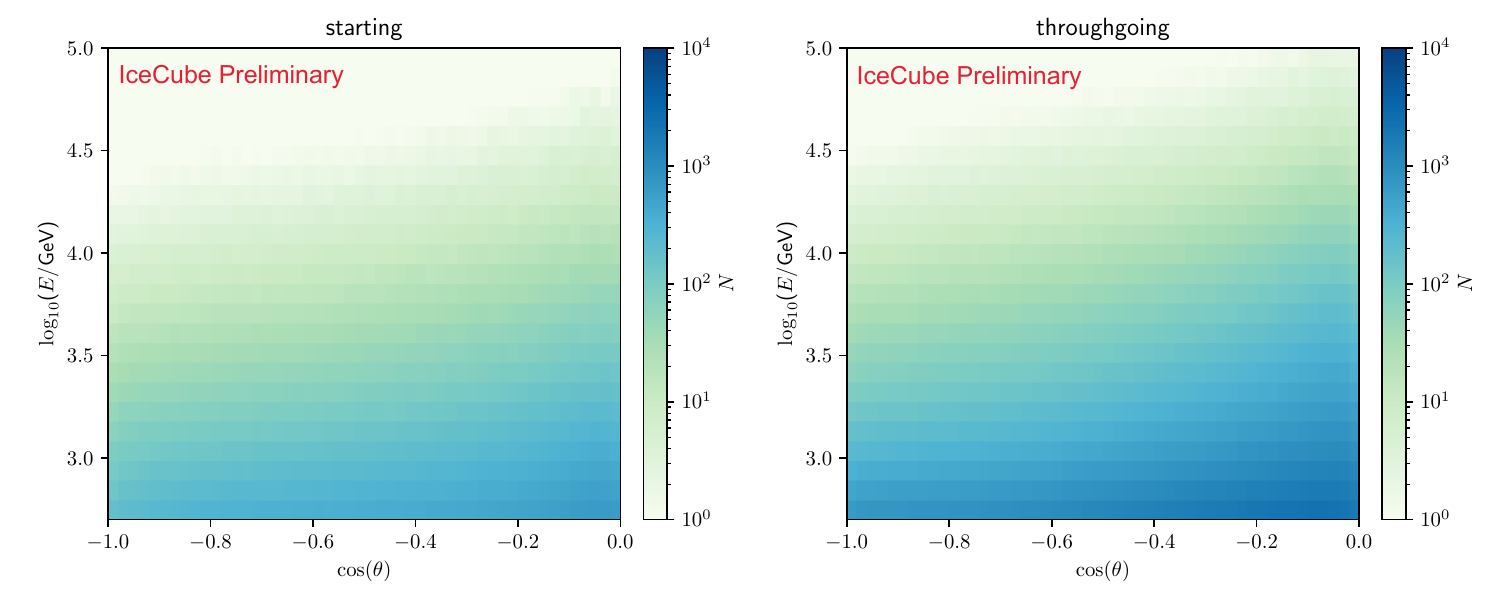}
\caption{The MC event distribution, weighted using nominal values of all model parameters, split into starting and throughgoing events. This is the total distribution from simulated conventional, astrophysical, and prompt neutrino fluxes attenuated through the PREM Earth density profile. The atmospheric fluxes are simulated using the Daemonflux construction~\cite{Yanez:2023lsy}.}\label{fig:event_dist}
\end{figure}

To perform the measurement, we perform a Poisson binned-likelihood analysis, in the Bayesian format.
The data and Monte Carlo (MC) events are split into 22 logarithmically-spaced bins of energy spanning 500 GeV to 100 TeV, and 50 uniformly-spaced bins in the range $\cos\theta \in [-1,0]$. 
We separately fit for starting and throughgoing distributions, which have distinct energy reconstruction performances.
In Fig.~\ref{fig:event_dist}, we show the nominal MC event distribution.
To construct the likelihood, we describe the MC event weights in terms of the Earth density parameters to be inferred as well as the nuisance parameters describing various sources of uncertainties and backgrounds. 
These stem from the conventional (from the decay of kaons and pions made in atmospheric cosmic-ray showers) and non-conventional fluxes (from astrophysical neutrinos, and from the prompt atmospheric decay of charmed hadrons), light propagation through the ice, and detection efficiencies.
To perform a Bayesian inference, we employ the \texttt{emcee} Markov Chain MC (MCMC) framework~\cite{emcee2013} to sample the posterior, which enables us to use the affine-invariant MCMC algorithm developed in Ref.~\cite{Goodman:2010dyf}.

\subsection{Earth parametrization}

\begin{figure}[t!]
\centering
\includegraphics[width=0.7\linewidth]{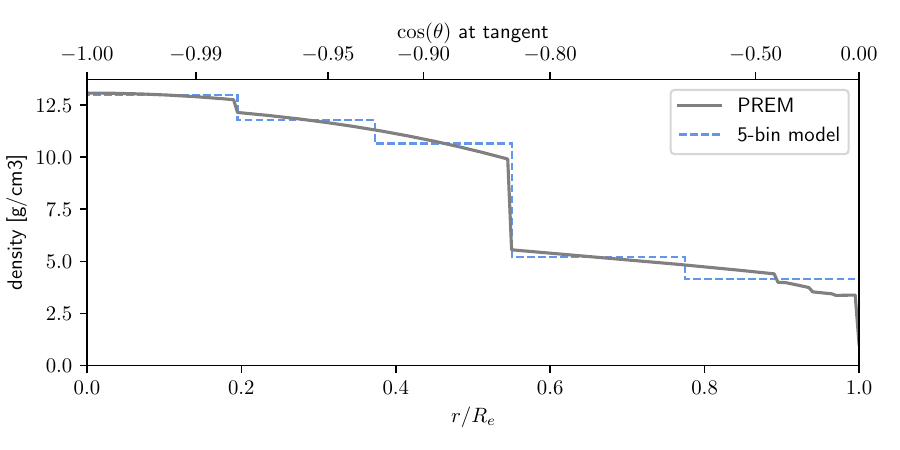}
\caption{The PREM density profile and the five-layer model. The x-axis plots the normalized radius of the Earth, where $R_e$ is the full-Earth radius. The region that is close to horizontal direction with respect to the IceCube location, in the range $\cos\theta \in [-0.1,0]$, is known as the horizon region. The top x-axis shows $\cos\theta$, where $\theta$ is the angle between the horizon and the tangent to the circle at the given radius.}\label{fig:binning_5B}
\end{figure}

\begin{figure}[t!]
\centering
\includegraphics[width=0.7\linewidth]{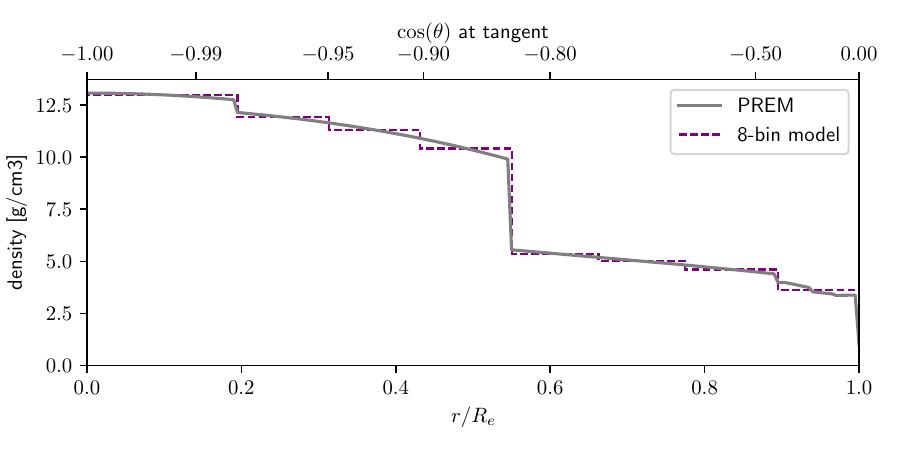}
\caption{The PREM density profile and the eight-layer model.}\label{fig:binning_8B}
\end{figure}

We evaluate two models of the earth profile. 
Primarily, we divide the radial profile into five constant-density layers, which is expected to yield the best precision on the density measurement, and offers a direction comparison with the work in Ref.~\cite{Donini:2018tsg}.
We also have a version with eight layers, to assess whether a more fine-grained inference is possible.
These profiles, in comparison to the PREM model, are shown in Figs.~\ref{fig:binning_5B} and~\ref{fig:binning_8B}. 
The parameters to infer are then the five or eight densities that describe the entire density profile. 

To parametrize the likelihood as a function of these densities, we pre-compute splines that return a transmittivity factor, $t$, (between 0 and 1) given a neutrino's true energy and traversed column density. 
We use \texttt{nuFATE}~\cite{Vincent:2017svp} to calculate this attenuation.
For each MC event, we calculate the column density, $\bar{x}(\theta)$, as a function of the Earth layer densities. 
We also calculate the column density $x(\theta)$ given the nominal density values, taken as the average PREM density in that layer -- this is the default column density used to generate the MC events.
The value of $t$ is then read from the spline given the event's column density and true energy, and the event weight is multiplied by a factor $\frac{t(\bar{x},E)}{t(x,E)}$.
This gives a way for us to quickly vary MC event weights and calculate the likelihood for each value of earth shell density.
In our Bayesian analysis, we employ uniform priors in linear space for the density value in each layer. 

\subsection{Systematic uncertainties and background}

The nuisance parameters that describe background fluxes and systematic uncertainties in this analysis are grouped into the following categories.
\begin{itemize}
    \item Detector parameters include an overall normalization and various parameters to describe the optical scattering/absorption of the glacial Antarctic ice as a function of depth, as well as the distinct ice in the IceCube bore holes that have been refrozen after detector construction.
    \item We use the Daemonflux~\cite{Yanez:2023lsy} construction to model the conventional atmospheric neutrino flux.
    We include the nuisance parameters from that model, which includes 10 parameters describing uncertainties on inclusive hadronic production cross sections at different energy points, and 6 parameters to model uncertainties on the cosmic ray flux. For details, refer to Ref.~\cite{Yanez:2023lsy}. We also include parameters that describe the atmospheric density and the rate of kaon energy loss, as well as an overall normalization.
    \item Non-conventional fluxes, meaning astrophysical and prompt atmospheric fluxes, are modelled by a broken power law flux model with two spectral indices, a spectral break, and a normalization.
    \item Finally, we have cross-section nuisance parameters which model the effect of an overall cross section uncertainty. 
\end{itemize}
There are 38 nuisance parameters in total.
All nuisance parameters have an independently-determined or estimated prior distribution, which are important inputs to the Bayesian analysis. 
For more details on the nuisance parameter treatment, refer to Ref.~\cite{IceCubeCollaboration:2024dxk}, which uses the same treatment. 

\begin{figure}[H]
\centering
\includegraphics[width=1.0\linewidth]{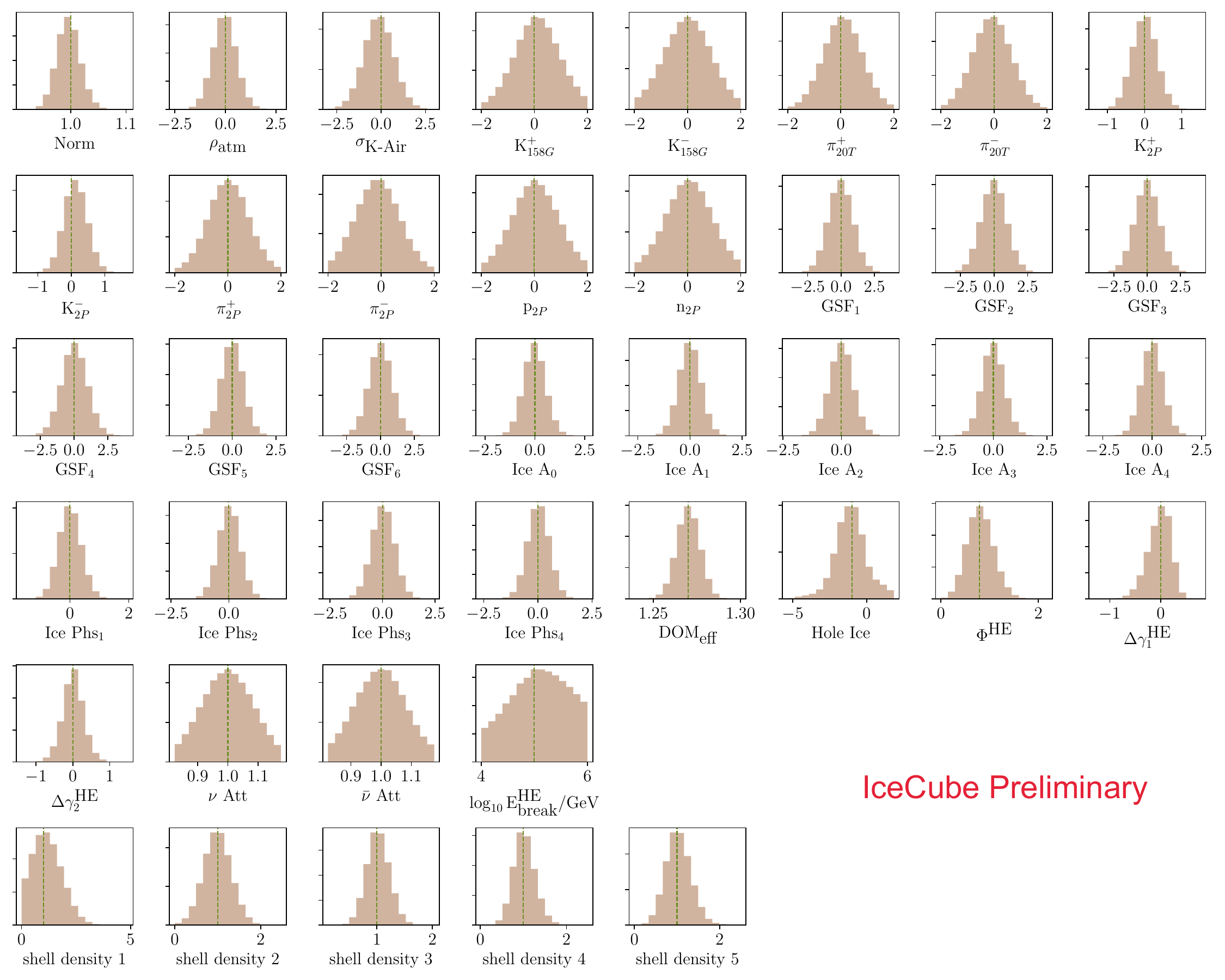}
\caption{The posterior distributions, which are derived from MCMC sampling, of all nuisance parameters. The density parameters are shown in the last row. The dashed vertical lines indicate the central prior values. For more details on the particular nuisance parameter names and what they mean, refer to Ref.\cite{IceCubeCollaboration:2024dxk}}
\label{fig:5B_grid}
\end{figure}

\section{Sensitivities}\label{sec:sensitivities}

\begin{figure}[t!]
\centering
\includegraphics[width=0.7\linewidth]{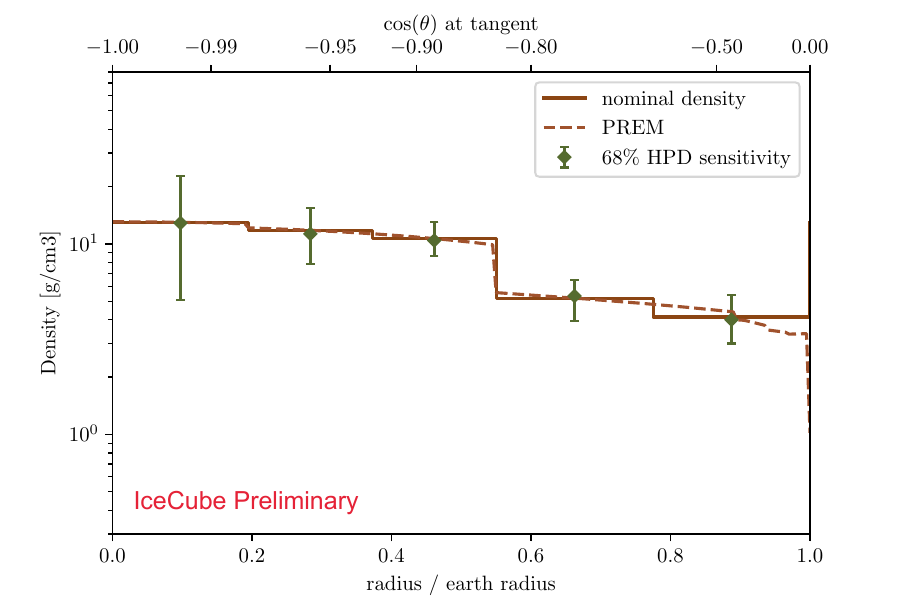}
\caption{The nominal five-layer density profile and the 68\% HPD regions for each layer overlaid. The PREM profile is also plotted for reference. As mentioned previously, the top horizontal axis shows the $\cos\theta$ of the direction that is tangent to the circle at the given radius.}\label{fig:5B_profile}
\end{figure}

\begin{figure}[t!]
\centering
\includegraphics[width=0.6\linewidth]{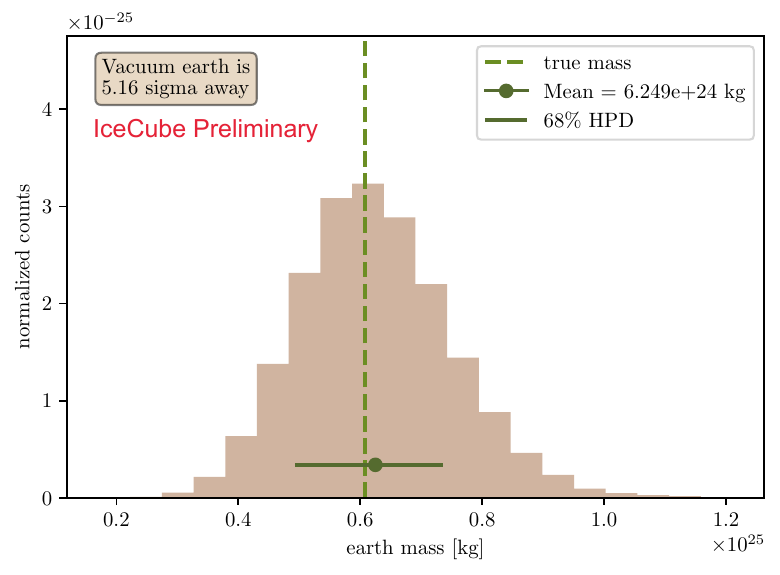}
\caption{The mass posterior distribution corresponding to the sensitivity. The posterior distribution is shown in brown, while the nominal mass is in the dashed line, and the mean and 68\% HPD of the posterior is shown in the horizontal bar.}\label{fig:5B_mass}
\end{figure}

\begin{figure}[t!]
\centering
\includegraphics[width=0.7\linewidth]{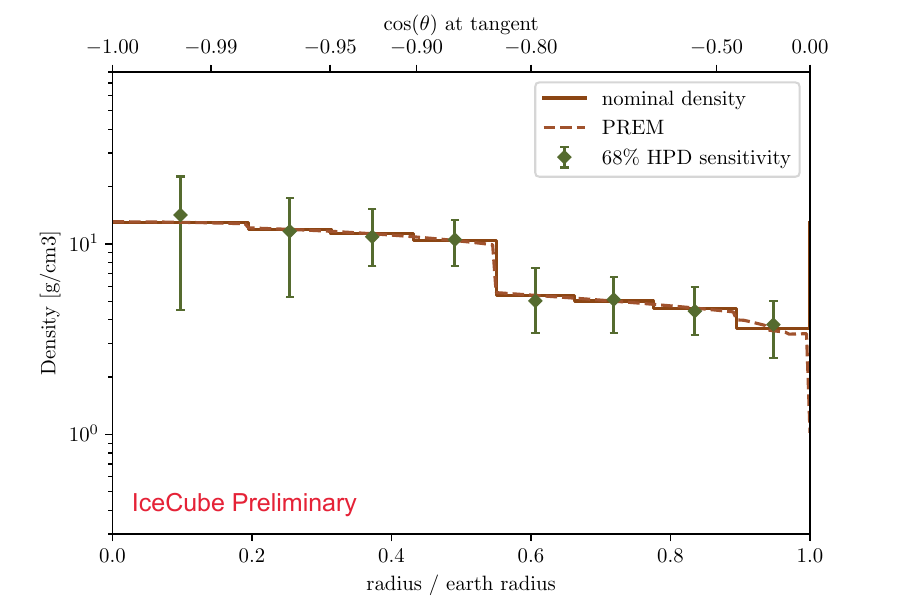}
\caption{The nominal eight-layer density profile and the 68\% HPD regions for each layer overlaid. The PREM profile is also plotted for reference.}\label{fig:8B_profile}
\end{figure}

We have computed sensitivities for the density parameters, where we fit to MC events held at their nominal prior values. 
This gives a good sense of how precise our measurement is expected to be.
In Fig.~\ref{fig:5B_grid} we show the posterior distributions for all nuisance and density parameters for the 5 layer model. 
In particular, the density parameter that is directly inferred is a scale factor on the average PREM density value in that layer.
We can compute the 68\% highest posterior density region (HPD) for each of the density parameters and overlay them on a density profile. This is shown in Fig.~\ref{fig:5B_profile}. 
The earth mass posterior corresponding to the density sensitivity is plotted in Fig.~\ref{fig:5B_mass}.
Finally, we also plot the sensitivity profile of the eight-layer model in Fig.~\ref{fig:8B_profile}.

\section{Conclusions}\label{sec:conclusions}

The sensitivities presented here correspond to the most precise measurement of the Earth using neutrino attenuation to date. 
Our Asimov sensitivities using both the five‐layer and eight‐layer parameterizations demonstrate that IceCube can probe the Earth’s radial density profile at a level that is more precise than previous neutrino measurements. 
In each layer, the 68\% HPD intervals encompass the corresponding nominal values, indicating a successful performance of the MCMC sampling while providing an estimate for the performance of a weak‐interaction–based measurement.
The posterior distributions of all nuisance and density parameters offer a natural way to estimate uncertainties in the measurement. 
The posterior for the total Earth mass likewise has uncertainties reduced substantially compared to the one‐year study in Ref.~\cite{Donini:2018tsg}, thanks to a combination of more data, improved energy reconstruction, and a more sophisticated nuisance parameter treatment. 
In summary, this work establishes a promising sensitivity for the upcoming measurement, and highlights the possibility of using neutrinos as a tomography tool in geophysics.

% Bibtex references:
\bibliographystyle{ICRC}
\bibliography{references}

\clearpage

\section*{Full Author List: IceCube Collaboration}

\scriptsize
\noindent
R. Abbasi$^{16}$,
M. Ackermann$^{63}$,
J. Adams$^{17}$,
S. K. Agarwalla$^{39,\: {\rm a}}$,
J. A. Aguilar$^{10}$,
M. Ahlers$^{21}$,
J.M. Alameddine$^{22}$,
S. Ali$^{35}$,
N. M. Amin$^{43}$,
K. Andeen$^{41}$
% ...
\\
\\
% ...
$^{10}$ Universit{\'e} Libre de Bruxelles, Science Faculty CP230, B-1050 Brussels, Belgium \\
% ...
$^{16}$ Department of Physics, Loyola University Chicago, Chicago, IL 60660, USA \\
$^{17}$ Dept. of Physics and Astronomy, University of Canterbury, Private Bag 4800, Christchurch, New Zealand \\
% ...
$^{21}$ Niels Bohr Institute, University of Copenhagen, DK-2100 Copenhagen, Denmark \\
$^{22}$ Dept. of Physics, TU Dortmund University, D-44221 Dortmund, Germany \\
% ...
$^{35}$ Dept. of Physics and Astronomy, University of Kansas, Lawrence, KS 66045, USA \\
% ...
$^{39}$ Dept. of Physics and Wisconsin IceCube Particle Astrophysics Center, University of Wisconsin{\textemdash}Madison, Madison, WI 53706, USA \\
% ...
$^{41}$ Department of Physics, Marquette University, Milwaukee, WI 53201, USA \\
% ...
$^{43}$ Bartol Research Institute and Dept. of Physics and Astronomy, University of Delaware, Newark, DE 19716, USA \\
% ...
$^{63}$ Deutsches Elektronen-Synchrotron DESY, Platanenallee 6, D-15738 Zeuthen, Germany \\
$^{\rm a}$ also at Institute of Physics, Sachivalaya Marg, Sainik School Post, Bhubaneswar 751005, India \\
% ...

\subsection*{Acknowledgments}

\noindent
The authors gratefully acknowledge the support from the following agencies and institutions:
USA {\textendash} U.S. National Science Foundation-Office of Polar Programs,
U.S. National Science Foundation-Physics Division,
U.S. National Science Foundation-EPSCoR,
U.S. National Science Foundation-Office of Advanced Cyberinfrastructure,
Wisconsin Alumni Research Foundation,
Center for High Throughput Computing (CHTC) at the University of Wisconsin{\textendash}Madison,
Open Science Grid (OSG),
Partnership to Advance Throughput Computing (PATh),
Advanced Cyberinfrastructure Coordination Ecosystem: Services {\&} Support (ACCESS),
Frontera and Ranch computing project at the Texas Advanced Computing Center,
U.S. Department of Energy-National Energy Research Scientific Computing Center,
Particle astrophysics research computing center at the University of Maryland,
Institute for Cyber-Enabled Research at Michigan State University,
Astroparticle physics computational facility at Marquette University,
NVIDIA Corporation,
and Google Cloud Platform;
Belgium {\textendash} Funds for Scientific Research (FRS-FNRS and FWO),
FWO Odysseus and Big Science programmes,
and Belgian Federal Science Policy Office (Belspo);
Germany {\textendash} Bundesministerium f{\"u}r Forschung, Technologie und Raumfahrt (BMFTR),
Deutsche Forschungsgemeinschaft (DFG),
Helmholtz Alliance for Astroparticle Physics (HAP),
Initiative and Networking Fund of the Helmholtz Association,
Deutsches Elektronen Synchrotron (DESY),
and High Performance Computing cluster of the RWTH Aachen;
Sweden {\textendash} Swedish Research Council,
Swedish Polar Research Secretariat,
Swedish National Infrastructure for Computing (SNIC),
and Knut and Alice Wallenberg Foundation;
European Union {\textendash} EGI Advanced Computing for research;
Australia {\textendash} Australian Research Council;
Canada {\textendash} Natural Sciences and Engineering Research Council of Canada,
Calcul Qu{\'e}bec, Compute Ontario, Canada Foundation for Innovation, WestGrid, and Digital Research Alliance of Canada;
Denmark {\textendash} Villum Fonden, Carlsberg Foundation, and European Commission;
New Zealand {\textendash} Marsden Fund;
Japan {\textendash} Japan Society for Promotion of Science (JSPS)
and Institute for Global Prominent Research (IGPR) of Chiba University;
Korea {\textendash} National Research Foundation of Korea (NRF);
Switzerland {\textendash} Swiss National Science Foundation (SNSF).

\end{document}